\documentclass[aps,prd,epsf,showpacs,amsmath,amssymb,10pt,twocolumn]{revtex4}
\usepackage{dcolumn} 
\usepackage{bm}
\usepackage{latexsym}
\usepackage{color,graphicx}

\newcommand{\e}{equation$\;$}

\newcommand{\be}{\begin{equation}}
\newcommand{\ee}{\end{equation}}
\newcommand{\ba}{\begin{eqnarray}}
\newcommand{\ea}{\end{eqnarray}}
\newcommand{\ban}{\begin{eqnarray*}}
\newcommand{\ean}{\end{eqnarray*}}

\newcommand{\n}[1]{\label{#1}}

\newcommand{\eq}[1]{(\ref{#1})}

\newcommand{\ph}{\ensuremath{{\phi}}}

\newcommand{\f}{\ensuremath{f(r)}}

\begin{document}

\title{Accelerated cosmic expansion in a scalar-field universe}

\author{$^1$Swastik Bhattacharya, $^1$Pankaj S. Joshi and $^2$Ken-ichi Nakao}
\email{swastik@tifr.res.in, psj@tifr.res.in, knakao@sci.osaka-cu.ac.jp}
\affiliation{$^1$Tata Institute for Fundamental Research, Homi Bhabha Road,
Mumbai 400005, India}
\affiliation{$^2$Department of Mathematics and Physics, Graduate School of 
Science, Osaka City University, Osaka 558-8585, Japan}

\begin{abstract} 
We consider here a spherically symmetric but inhomogeneous universe filled with a massless scalar field. The model obeys two constraints. The first one is that the gradient of the scalar field is timelike everywhere. The second constraint is that the radial coordinate basis vector is a unit vector field in the comoving coordinate system. We find that the resultant dynamical solutions compose a one-parameter family of self-similar models which is known as the Roberts solution. The solutions are divided into three classes. The first class consists of solutions with only one spacelike singularity in the synchronous-comoving chart. The second class consists of solutions with two singularities which are null and spacelike, respectively. The third class consists of solutions with two spacelike singularities which correspond to the big bang and big crunch, respectively. We see that, in the first case, a comoving volume exponentially expands as in an inflationary period; the fluid elements are accelerated outwards form the symmetry center, even though the strong energy condition is satisfied. This behavior is very different from that observed in the homogeneous and isotropic universe in which the fluid elements would move outwards with deceleration, if the strong energy conditions are satisfied. We are thus able to achieve the accelerated expansion of the universe for the models considered here, without a need to violate the energy conditions. The cosmological features of the models are examined in some detail.

\end{abstract}
\maketitle

\section{Introduction}

The big bang universe model has succeeded to explain three prominent 
observational facts, {\it i.e.}, the Hubble's law, the abundance of
light elements, and the 2.7K cosmic microwave background radiation
(CMBR)
\cite{Weinberg}. 
Further, within this model, the formations of various 
structures in the universe, i.e., stars, galaxies, clusters of galaxies, etc, seem 
to be naturally explained if we assume the existence of dark matters
\cite{Peebles}.  
In the original big bang model, there are three basic unnatural aspects,
which are known as the horizon problem, flatness problem, and the 
monopole problem. Now 
it is widely believed that these three 
problems are the evidences for existence of the so called 
inflationary period in the past of our universe
\cite{Sato,Guth,nde}. 
In this period, by virtue of the accelerated cosmic volume expansion, 
initial inhomogeneities are stretched, the space is made 
sufficiently flat, and at the same time, moderate quantum instabilities generate 
the seeds of the present structures in the universe. The theoretical prediction 
of the inflationary scenario basically agrees well with the WMAP data on 
CMBR\cite{WMAP}. 

Recent observational results also imply that our universe has again 
entered into another inflationary period at about $z\sim0.3$. This is known 
as the dark energy problem. 
The WMAP data and the observations of the distance-redshift relations of 
supernovae\cite{Riess-1,Perlmutter,Knop,Riess-2}  
mean that the speed of the cosmic volume expansion is necessarily accelerated 
if the universe is homogeneous and isotropic. Further, if general relativity 
is the correct theory of gravity, then 
the accelerated cosmic expansion implies the existence of a dark energy 
component which does not satisfy the strong energy condition, i.e., effectively 
it has to have an 
equation of state given as $P=w\rho$ with $w<-1/3$. 
It is far from clear today what such a dark energy component could be.
The simplest answer to this conundrum could be the existence of a non-vanishing 
cosmological constant in the universe, since in that case, only one constant 
can explain all of the observational data. However, it does not at all appear to be 
easy to explain the origin of such a cosmological term. 
Therefore, the issue is under an active investigation and several approaches 
to resolve the problem are being pursued simultaneously. These include the
existence of a quintessence field, using an alternative theory of gravity
(see, for example Ref.\cite{Caldwell}), and also inhomogeneous universe 
models, as we shall discuss below.

In this paper, we consider a self-similar model for the spherically
symmetric but inhomogeneous universe, filled with a massless scalar
field which is minimally coupled to gravity. 
This solution to Einstein equations has been discussed in the literature 
by Roberts \cite{Roberts} ,
Brady
\cite{Brady}  
and Oshiro, Nakamura and Tomimatsu \cite{Oshiro}. 
However, these authors used this solution only to study the 
gravitational collapse of a massless scalar field and used it as a 
model for the collapsing matter. In this paper we will focus on the 
cosmological significance of this solution. All of the physically reasonable 
energy conditions hold for this model universe. Although this solution 
does not directly relate to our present universe, there are certain
interesting properties that the models exhibit as we shall point out here. 
In the region where the gradient of the scalar field is timelike, the
system is equivalent to 
a stiff perfect fluid, given by the equation of state $P=\rho$, 
and we can consider motions of the fluid elements. As we show here,
in this solution the fluid elements are accelerated outwards, 
even though the strong energy condition holds.

Here we should mention the models for a dust-filled inhomogeneous 
universe, which have been considered to resolve the dark energy problem 
without introducing any exotic matter-components, or without invoking 
any alternative theories of gravity. 
Mainly two approaches have been used in this connection.
The first one invokes  
the backreaction effect of inhomogeneities in the universe, which causes 
the acceleration of the cosmic volume expansion in an average 
sense\cite{Futamase,Russ,Nambu,Kozaki,Kolb,Nambu-Tanimoto,Kai}.
In the other approach,  one invokes a spherically symmetric but 
inhomogeneous velocity field 
which causes the apparent acceleration of the cosmic volume 
expansion\cite{Zehavi,Tomita,Celerier,Iguchi,Vandervelt,Alnes,Garcia}. 
At present, it seems to be difficult to resolve the dark energy problem using 
the first approach (see e.g. \cite{Flanagan,Hirata,Ishibashi,Singh}
for a discussion on this issue), 
and also there are some arguments which discuss the viability or
otherwise of the second approach
\cite{Zibin,Garcia-2}. 

We would like to emphasize that the present solution 
that we discuss and consider here is different from both of the above 
possible alternatives that have been used for dust-filled inhomogeneous 
universe models. In the present solution, 
each fluid element is accelerated outwards in the real and exact 
sense from the observer at the center of symmetry.

The paper is organized as follows. In Sec.II, we briefly review 
the massless scalar field minimally coupled to gravity, 
and then, in Sec.III, the basic equations in a comoving reference frame 
are presented. In Sec.IV, by imposing a constraint 
that the radial coordinate basis vector 
is a unit vector field, we obtain the solution. We believe our
method gives some insights into the properties of the solution
we derived this way. In Sec.V, we show that the 
charts of the solutions obtained in some of the classes
we considered do not cover whole of the spacetime.
Therefore, we perform two kinds of extensions respectively, 
in Sec.VI and Sec.VII. In Sec.VI, we also discuss and analyze the 
global structure of the solutions. Finally, Sec.VIII is devoted to 
a summary and discussion. 

In this paper, we adopt the abstract index notation\cite{Wald}; 
the Latin indices denote the type of the tensor, whereas the Greek 
indices mean the components 
with respect to the coordinate basis. We use the units $8\pi G=c=1$.

\section{Massless scalar field}

The massless scalar field $\ph(x^a)$ minimally coupled to gravity on a spacetime  
$(M, g_{ab})$ is described by the Lagrangian, 
\be
{\cal L}=-\frac{1}{2}\ph_{;a}\ph_{;b}g^{ab}.
\n{lag}
\ee
The corresponding Euler-Lagrange equation is 
$\ph_{;ab}g^{ab}=0$,
and the stress-energy tensor for the scalar field, as calculated 
from this Lagrangian is, 
\be
T_{ab}=\ph_{;a}\ph_{;b}-\frac{1}{2}g_{ab}\left(\ph_{;c}\ph_{;d}
g^{cd}\right).
\n{emt}
\ee
In case that the gradient of the massless scalar field 
$\phi_{;a}$ is timelike or spacelike, we can introduce a unit vector field
\be
u_a=\frac{\phi_{;a}}{N},
\ee
where $N=\sqrt{|\phi^{;b}\phi_{;b}|}$. Then the stress-energy tensor 
of the massless scalar field is rewritten in the form
\be
T_{ab}=\frac{N^2}{2}\left(2u_a u_b+\varepsilon g_{ab}\right), \label{st-tensor}
\ee
where $\varepsilon=-1$ for spacelike $\phi_{;a}$, while $\varepsilon=+1$ 
for timelike $\phi_{;a}$. Clearly, $u^a$ is one of eigenvectors of $T_{ab}$. 
In this case, the massless scalar field is categorized into a {\it Type I}
matter field\cite{haw},  
{\it i.e.}, it admits one timelike and three spacelike eigenvectors.
So, at each point $q\in M$, we can then express the tensor $T^{ab}$ 
in terms of an orthonormal basis $(E_0^a,E_1^a,E_2^a,E_3^a)$, 
where $E_0^a$ is a timelike eigenvector with an eigenvalue $\rho$ 
and $E_{A}^a$ $(A=1,2,3)$ are three spacelike 
eigenvectors with eigenvalues $p_A$. Here $\rho$ represents 
the energy density of the scalar field as measured by an observer 
with a 4-velocity $E_0^a$ at $q$, 
and the eigenvalues $p_A$ represent the principal pressures 
in three spacelike directions $E_A^a$. 

In the case of spacelike $\phi_{;a}$, we can put $E_1^a=u^a$. 
Then the stress-energy tensor \eq{st-tensor} is written in the form
\begin{equation}
T^{ab}=\frac{N^2}{2}\left(E_0^a E_0^b+E_1^a E_1^b -E_2^a E_2^b 
-E_3^a E_3^b\right).
\end{equation}
It is seen from the above form that $\rho=p_1=N^2/2\geq 0$, whereas 
$p_2=p_3=-N^2/2\leq0$. 

In the case of timelike $\phi_{;a}$, we can put $E_0^a=u^a$. 
Then the stress-energy tensor \eq{st-tensor} is written in the form
\begin{equation}
T^{ab}=\frac{N^2}{2}\left(E_0^a E_0^b+E_1^a E_1^b +E_2^a E_2^b 
+E_3^a E_3^b\right).
\end{equation}
It is seen from the above form that $\rho=p_1=p_2=p_3=N^2/2\geq 0$. This is 
equivalent to the stress-energy tensor of the stiff perfect fluid. 

In the case that $\phi_{;a}$ is null, the stress-energy tensor takes the 
following form,
\be
T_{ab}=\sigma k_a k_b,
\ee
where $k_a$ is a null vector field and $\sigma$ is a non-negative function. 
This is equivalent to the stress-energy tensor 
of the null dust, and is categorized into a {\it Type II} matter field.

\section{Basic equations}

Hereafter, we focus on the spherically symmetric system.

\subsection{Comoving coordinate system}

If the gradient of the scalar field is not null, as mentioned 
in the previous section, the stress-energy tensor has four eigenvectors 
$(E_0^a,E_A^a)$. In this case, we can choose the spherically symmetric 
coordinates $(t,r,\theta,\varphi)$ such that the coordinate basis vectors 
are parallel to these eigenvectors. This is the {\it comoving} coordinate 
system in the sense that there is no energy flux, i.e., the time-space 
components of the stress energy-tensor are vanishing. The line element 
in this coordinate system is then written in the following form,
\begin{equation}
ds^2=-e^{2\nu(t,r)}dt^2+e^{2\psi(t,r)}dr^2+R^2(t,r)d\Omega^2, \label{metric}
\end{equation}
where $d\Omega^2$ is the line element on a unit two-sphere. It is noted that 
there are still two scaling freedoms of one variable left in $t$ and $r$. 
In the models considered here, $R$ is a monotonically increasing 
function of $r$. 

In the spherically symmetric system, a non-trivial time-space 
component is 
\begin{equation}
T_{tr}=\dot{\phi}\phi',
\end{equation}
where $(\dot{~})$ denotes the partial derivative with respect to $t$ and 
$(')$ with respect to $r$. Thus, in the comoving system, one of the  
$\dot{\phi}$ and $\phi'$ must vanish. 

Hereafter, as mentioned in Sec.I, we assume that $\phi_{;a}$ is 
everywhere timelike, i.e., $\phi=\phi(t)$, and thus this system 
is equivalent to that of the stiff perfect fluid. 
Here it is worth noting that if the norm of $\phi_{;a}$ does not 
have a definite sign, the comoving coordinate system cannot cover 
whole of the spacetime. 

The components of the stress-energy tensor are then
given by,
\begin{equation}
T^\mu{}_\nu={\rm diag}[-\rho,\rho,\rho,\rho]\;\;,
\label{eq:em}
\end{equation}
where
\begin{equation}
\rho=\frac{1}{2}e^{-2\nu}\dot{\ph}^2. 
\end{equation}

\subsection{Einstein equations}

The dynamic evolution of the initial data, as specified on a 
spacelike surface of constant time is determined by 
the Einstein equations. For the metric (\ref{metric}), 
using the definitions 
\begin{eqnarray}
G(t,r)&=&e^{-2\psi}(R^{\prime})^{2},  \label{eq:G-def}\\
H(t,r)&=&e^{-2\nu} (\dot{R})^{2}, \label{eq:H-def}
\end{eqnarray}
and
\begin{equation}
F=R(1-G+H),
\label{eq:ein4}
\end{equation}
the independent Einstein equations for the 
massless scalar field are,
\begin{equation}
F^{\prime}= \frac{1}{2}e^{-2\nu}\dot{\ph}^2R^{2}R^{\prime}\;, 
\label{eq:ein1}
\end{equation}
\be
\dot{F}=-\frac{1}{2}e^{-2\nu}\dot{\ph}^2R^{2}\dot{R} \;,
\n{eq:ein1a}
\ee
\begin{equation}
\partial_t\left(R^2e^{\psi-\nu}\dot{\ph}\right)=0\;,
\label{eq:ein2}
\end{equation}
\begin{equation}
-2\dot{R}'+R'\frac{\dot{G}}{G}+\dot{R}\frac{H'}{H}=0\; .
\label{eq:ein3}
\end{equation}
Here the function $F=F(t,r)$ has an interpretation of the mass 
function for the collapsing cloud, and it gives the total mass in a 
shell of comoving radius $r$ on any spacelike slice $t=const$.

\section{Solving the Einstein equations}

The function $R(t,r)$ is the area radius of a shell labeled 
`$r$' at an epoch `$t$'. For the sake of definiteness, let us consider 
the situation of an expanding universe, so we have $\dot R>0$ because
we are considering here the expanding branch of the solutions. If $\dot R$ 
changes sign then that corresponds to the recollapse of the 
field during the dynamical evolution of the universe. 
We note that \e \eq{eq:ein2} is the Klein-Gordon equation for the scalar 
field, which is a part here of the Einstein equations via the Bianchi 
identities. We can integrate this equation to get
\begin{equation}
R^2e^{\psi-\nu}\dot{\ph}=r^2\f,
\n{kg}
\end{equation}
where $\f$ is an arbitrary function of integration.
Solution of these equations, subject to the initial data and energy 
conditions, would determine the time evolution of the system.

We now construct a class of solutions with $\psi=0$. 
This is the second assumption that we make here, namely that the 
radial coordinate basis vector is an unit vector field. From \eqref{eq:ein3}, we have
\begin{equation}
H=H(t) \label{eq:H-sol}
\end{equation}
Also, using Eqs.~\eqref{kg} and \eqref{eq:H-sol}, Eq.~\eqref{eq:ein1a} 
can be integrated to give 
\begin{equation}
F= \frac{1}{2} \frac{r^4f^2(r)}{R}+h(r), \label{eq:F-sol}
\end{equation}
where $h(r)$ is an arbitrary function of integration. Together with the above 
equation, Eq.~(\ref{eq:ein1}) leads
\begin{equation}
\left(\frac{1}{R}\right)'+\frac{(r^2f)'}{r^2f}\frac{1}{R}+\frac{h'}{r^4f^2}=0.
\end{equation}
The solution of the above equation is given by 
\begin{equation}
R= \frac{r^2f(r)}{g(t)+p(r)}, \label{eq:R-sol}
\end{equation}
where $g(t)$ is an arbitrary function of integration, and
\begin{equation}
p(r)=-\int\frac{h'(r)}{r^2f(r)}dr. \label{eq:p-def}
\end{equation}
Here, by a scaling freedom of $t$, we set 
\begin{equation}
\phi(t)=\varphi t,
\end{equation}
where $\varphi$ is a constant with a dimension of length-inverse (note that 
the scalar field is dimensionless in our unit). Then we have 
$\varphi e^{-\nu}R^2=r^2f(r)$ from Eq.~(\ref{kg}). Further, using Eqs.~\eq{eq:H-def} 
and \eq{eq:H-sol}, we have
\begin{equation}
\left(\frac{dR^{-1}}{dt}\right)^2=\frac{\varphi^2H(t)}{r^4 f^2(r)}.
\end{equation}
Using Eq.~(\ref{eq:R-sol}), the above equation implies
\begin{equation}
\dot{g}^2(t)=\varphi^2H(t). \label{eq:g-dot}
\end{equation}

Substituting the above results into Eq.~\eqref{eq:ein4}, we have
\begin{equation}
H+1=\frac{1}{2}(g+p)^2+\frac{h}{r^2f}(g+p) 
+\left[\frac{(r^2f)'(g+p)+h'}{(g+p)^2}\right]^2.
\label{eq:ein4-2}
\end{equation}
The left hand side of the above equation depends on only $t$, 
and thus the right hand side should also depend on only $t$. This equation gives 
a constraint  on the functions $f$ and $h$. 

As long as we focus on the dynamical situations, 
we can see from Eq.~(\ref{eq:ein4-2}) that the most general 
allowed form of $r^2f(r)$ is 
\begin{equation}
rf(r)=\alpha,
\end{equation}
where $\alpha$ is a positive dimensionless constant, and further 
$h=0$ (see Appendix \ref{ApA}). 
Then, Eq.~\eqref{eq:ein4-2} leads to 
\begin{equation}
\varphi^{-1}\dot{g}(t)=\pm \left[\frac{1}{2}g^2(t)+\frac{\alpha^2}{g^2(t)}
-1\right]^{\frac{1}{2}}. \label {dn2}
\end{equation}
We take the negative sign in the equation so that $\dot{R}>0$. 
This can be easily solved, and we have, 
\begin{equation}
g=\sqrt{
1+\frac{1}{2}\left(Ce^{-\sqrt{2}\varphi t}+
\frac{1-2\alpha^2}{Ce^{-\sqrt{2}\varphi t}}\right)
},
\end{equation}
where $C$ is an integration constant. 
So the metric we get is  
\begin{equation}
ds^2= -\frac{(\varphi\alpha r)^2}{g^4(t)}dt^2+dr^2+\frac{(\alpha r)^2}{g^2(t)}d\Omega^2
\label{eq:metric-2}
\end{equation}

Also, we have 
\begin{equation}
F= \frac{1}{2}\alpha rg(t)~~~~{\rm and}~~~~\rho=\frac{1}{2}\frac{g^4(t)}{(\alpha r)^2}.
\label{eq:density}
\end{equation}
From Eq.~(\ref{eq:density}), we see that the energy density $\rho$ diverges 
at $r=0$ if $g$ does not vanishes. Moreover, even for $r>0$, $\rho$ diverges 
if $g$ diverges. Since the Ricci scalar is proportional to 
$\rho$, $g=\infty$ corresponds to the 
scalar polynomial singularity\cite{haw}. 

Here, we choose $C$ so that $C=\sqrt{2\alpha^2-1}$ for $\alpha^2>1/2$, 
$C=1$ for $\alpha^2=1/2$, and $C=\sqrt{1-2\alpha^2}$ for $\alpha^2<1/2$. 
Then, we have following three distinct classes of solutions.

\subsection{The case of $\alpha^2>1/2$}

In this case, we obtain
\begin{equation}
g(t)= \left[1-\sqrt{2\alpha^2-1}\sinh\left(\sqrt{2}\varphi t\right)\right]^{\frac{1}{2}}.  
\end{equation}
The domain of time $t$ is $-\infty<t<t_{\rm b}$, where
\begin{equation}
t_{\rm b}:=\frac{1}{2\sqrt{2}\varphi}\ln\left|\frac{\sqrt{2}\alpha+1}{\sqrt{2}\alpha-1}\right|.
\end{equation}
For $t\rightarrow-\infty$, $g$ diverges, and thus the energy density 
and the Ricci scalar also diverge for $r>0$. 
As will be shown in the next section, $t\rightarrow -\infty$ 
is not infinity, and hence 
this should be regarded as an initial singularity (big bang). 
At $t=t_{\rm b}$, $g$ vanishes, and thus any 2-dimensional 
spheres with positive comoving radii have infinite area, since 
the area $A$ with a comoving radius $r$ 
is given by 
\begin{equation}
A=4\pi(\alpha r)^2/g^2. 
\end{equation}
As will be shown later, $t=t_{\rm b}$ is on the future null infinity 
except at $r=0$;  
a ``point'' $(t,r)=(t_{\rm b},0)$ is a sphere with finite area. 
As will be shown later, the extension over this sphere is possible.

\subsection{The case of $\alpha^2=1/2$}

In this case, we obtain
\begin{equation}
g(t)= \left[1+\exp\left(-\sqrt{2}\varphi t\right)\right]^{\frac{1}{2}} .
\end{equation}
The domain of time is $-\infty<t<\infty$. 
As in the case of $\alpha^2>1/2$, $t=-\infty$ 
corresponds to an initial singularity at which $g$ diverges. 
However, in contrast with the case of 
$\alpha^2>1/2$, $g$ is always larger than unity and finite except on $t=-\infty$. 
This chart is inextendible.

\subsection{The case of $\alpha^2<1/2$}

In this case, we obtain
\begin{equation}
g(t)= \left[1+\sqrt{1-2\alpha^2}\cosh\left(\sqrt{2}\varphi t\right)\right]^{\frac{1}{2}}.  
\end{equation}
The domain of time is also $-\infty<t<\infty$. 
As in the previous cases of $\alpha^2 \geq 1/2$, $t\rightarrow-\infty$ 
corresponds to an initial singularity at which $g$ diverges. 
However, in contrast with the  
case of $\alpha^2 = 1/2$, $g$ diverges also 
in the limit of $t\rightarrow \infty$. 
As will be shown later, the limit of $t\rightarrow\infty$ 
is not infinity, and thus this is 
a final singularity (big crunch). This chart is also inextendible.

\section{Some properties of  solutions}

In this section, we shall analyze a few properties of this solution,
including its extendibility. For this purpose, we introduce a new time 
coordinate $\tau$ defined by 
\begin{equation}
d\tau=\frac{dt}{g^2(t)}. \label{eq:dtau}
\end{equation}
The line element with this new time coordinate is then given by
\begin{equation}
ds^2=-(\varphi\alpha r)^2d\tau^2+dr^2+\frac{(\alpha r)^2}{g^2}d\Omega^2.
\label{eq:metric-21}
\end{equation}

This new time coordinate is proportional to the 
proper time for a  comoving observer along $r=$constant line. Thus, this 
is a geometrically and physically meaningful quantity. 
We discuss the cases of $\alpha^2>1/2$, $\alpha^2=1/2$, 
and $\alpha^2<1/2$, separately.

\subsection{The case of $\alpha^2>1/2$}
By integration of Eq.~(\ref{eq:dtau}) with an appropriate integration constant, we have
\begin{equation}
\tau=\frac{1}{2\alpha\varphi}
\ln\left| 
\frac{1+\sqrt{(\sqrt{2}\alpha+1)/(\sqrt{2}\alpha-1)}e^{\sqrt{2}\varphi t}}
{1-\sqrt{(\sqrt{2}\alpha-1)/(\sqrt{2}\alpha+1)}e^{\sqrt{2}\varphi t}}
\right|  .
\end{equation}
We can easily see from the above equation that $\tau$ vanishes for $t\rightarrow-\infty$. 
This means that $t\rightarrow-\infty$ is not infinity. As mentioned, since the energy density $\rho$ 
and Ricci scalar diverge for $t\rightarrow-\infty$, this is the initial singularity. 
By contrast, $\tau$ becomes infinite in the limit of $t\rightarrow t_{\rm b}$. 
As will be shown later, $t=t_{\rm b}$ corresponds to the future null infinity.

\subsection{The case of $\alpha^2=1/2$}

In this case, we obtain
\begin{equation}
\tau=\frac{1}{\sqrt{2}\varphi}
\ln\left| e^{\sqrt{2}\varphi t}+1
\right|. 
\end{equation}
We can easily see that $\tau$ vanishes in the limit of $t\rightarrow -\infty$, whereas
$\tau$ becomes infinite in the limit of $t\rightarrow\infty$. Thus, $t\rightarrow-\infty$ is 
the initial singularity also in this case, whereas, as will be shown later, $t\rightarrow\infty$ 
corresponds to the future timelike infinity.

\subsection{The case of $\alpha^2<1/2$}

In this case, we have
\begin{equation}
\tau=\frac{1}{2\alpha\varphi}
\ln\left| 
\frac{1+\sqrt{(1+\sqrt{2}\alpha)/(1-\sqrt{2}\alpha)}e^{\sqrt{2}\varphi t}}
{1+\sqrt{(1-\sqrt{2}\alpha)/(1+\sqrt{2}\alpha)}e^{\sqrt{2}\varphi t}}
\right|.  
\end{equation}
Also in this case, $\tau$ vanishes in the limit of $t\rightarrow-\infty$, and this ``moment'' 
corresponds to the initial singularity. In contrast to the case of $\alpha^2=1/2$, 
we have, for $t\rightarrow \infty$, 
\begin{equation}
\tau\longrightarrow \tau_{\rm c}:=\frac{1}{2\alpha\varphi}
\ln\left|\frac{1+\sqrt{2}\alpha}{1-\sqrt{2}\alpha}\right|. 
\end{equation}
Thus, the limit of $t\rightarrow\infty$ does not corresponds to an infinity, but 
the final singularity. 

Using the above results, we obtain
\begin{equation}
g^2=\frac{4\alpha^2 e^{2\alpha\varphi\tau}}
{(\sqrt{2}\alpha-1)e^{4\alpha\varphi\tau}+2e^{2\alpha\varphi\tau}-\sqrt{2}\alpha-1},
\label{eq:g-sol}
\end{equation}
for arbitrary positive $\alpha$. 
If the denominator in the right hand side of Eq.~(\ref{eq:g-sol}) vanishes, then $g$ diverges. 
The denominator vanishes if 
\begin{equation}
e^{2\alpha\varphi\tau}=1
\end{equation}
or
\begin{equation}
e^{2\alpha\varphi\tau}=\frac{1+\sqrt{2}\alpha}{1-\sqrt{2}\alpha} 
\end{equation}
is satisfied. The second root is meaningful only if $\alpha^2<1/2$, since $\tau$ 
should be a real number. 
The first root is $\tau=0$ which corresponds to the big bang, whereas the 
second one is $\tau=\tau_{\rm c}$ which corresponds to the big crunch. 

It might be a remarkable fact that, in the case of $\alpha^2>1/2$, 
the areal radius $R$ behaves as 
\begin{equation}
R^2\longrightarrow \frac{\sqrt{2}\alpha -1}{4\alpha^2}r^2e^{2\alpha\varphi\tau},
\end{equation}
for $\tau \gg (\alpha\varphi)^{-1}$.
Thus it is seen that the comoving volume exponentially expands as in 
the inflationary period;  
and that the acceleration of the cosmic volume expansion is realized, 
even though the strong energy condition holds. 

The world interval of the submanifold $(\tau,r)$ in Eq.~(\ref{eq:metric-21}) 
takes the same form as that of the Rindler spacetime. 
Since the Rindler spacetime is extendible, we are led to infer that
this solution might also be extendible. 
We discuss the maximal extension of this solution 
in the following two sections. 

\section{Analytic Extension}

Following the prescription of the maximal extension for the Rindler spacetime, 
we introduce following new coordinates 
\begin{equation}
T=r\sinh(\varphi\alpha\tau)~~~~~{\rm and}~~~~~X=r\cosh(\varphi\alpha\tau), \label{eq:TX-def}
\end{equation}
then we have 
\begin{equation}
ds^2=-dT^2+dX^2+R^2d\Omega^2. \label{eq:metric-3}
\end{equation}
From Eq.~(\ref{eq:TX-def}), we have
\begin{equation}
X^2-T^2=r^2~~~~~~{\rm and}~~~~~~\frac{T}{X}=\tanh(\varphi\alpha\tau).
\end{equation}
The first equation implies that $r$-constant curves are timelike 
hyperbolic curves, whereas $\tau$-constant curves are spacelike 
straight lines, in the $(T,X)$ plane. 
The singularity $\tau=0$ corresponds to $T=0$, whereas 
$r=0$ is null $T=\pm X$, and thus $T/X=\pm1$. This implies that 
$r=0$ also corresponds to $\tau=\pm\infty$ except at $T=X=0$. 

The square of the areal radius $R^2$ is written as a function of 
$T$ and $X$ in the form 
\begin{equation}
R^2=T(\sqrt{2}\alpha X-T). \label{eq:TX-R}
\end{equation}
The above equation explicitly shows the self-similarity 
of the present solution, since all the dimensionless variables are 
written as functions of a self-similar variable $\xi:=X/T$. 
Further, there is only one parameter $\alpha$. Thus, the solutions 
obtained here compose a one-parameter family of self-similar solutions 
which is known as the Roberts solution\cite{Roberts}. 

In order that $R^2$ is non-negative, both of  
$T\geq0$ and $T\leq\sqrt{2}\alpha X$ have to be satisfied, or both of 
$T\leq0$ and $T\geq\sqrt{2}\alpha X$ have to be satisfied. 
The boundary of these two regions in $(T,X)$ plane 
is the singularity $T=0$ ($\tau=0$). Thus 
we cannot regard these regions in $(T,X)$ as two regions of 
one spacetime manifold: each region should be regarded as an 
independent spacetime manifold. 
Here the former, $T\geq0$ and $T\leq\sqrt{2}\alpha X$,
is of our interest. 

The energy density $\rho$ is written in the form
\begin{equation}
\rho=\frac{\alpha^2(X^2-T^2)}{2T^2(\sqrt{2}\alpha X-T)^2}. \label{eq:TX-rho}
\end{equation}
Thus the spacetime singularities are located 
at $T=0$ and $T=\sqrt{2}\alpha X$. 

It is useful for understanding the physical situation of this 
spacetime to 
examine the expansions of future-directed null, and also to 
consider the Misner-Sharp mass\cite{Misner} 
(see Appendix \ref{ApMS}). 
In the coordinate system (\ref{eq:metric-3}), 
the expansions of outgoing and ingoing null 
are given, respectively, by
\begin{eqnarray}
\vartheta_{+}&=&\frac{1}{R^2}\left[\alpha X-(\sqrt{2}-\alpha)T\right], 
\label{eq:out-expansion}\\
\vartheta_{-}&=&\frac{1}{R^2}\left[\alpha X-(\sqrt{2}+\alpha)T\right].
\label{eq:in-expansion}
\end{eqnarray}
In accordance with Hayward's definition\cite{Hayward}, 
a region or a surface is said to be {\it trapped} if $\vartheta_{+}\vartheta_{-}>0$, 
{\it marginal} if $\vartheta_{+}\vartheta_{-}=0$, {\it untrapped} if 
$\vartheta_{+}\vartheta_{-}<0$. A trapped region or a trapped surface of $\vartheta_{-}>0$ is 
said to be past trapped, whereas a trapped region or a trapped surface of 
$\vartheta_{+}<0$ is said to be future trapped. 
Similarly, a region or surface of $\vartheta_{-}=0$  
is said to be past marginal, whereas a region or surface of $\vartheta_{+}=0$ 
is said to be future marginal. 

From Eq.~(\ref{mass-def}), we have the Misner-Sharp mass $M_{\rm MS}$ of 
Eq.~(\ref{eq:metric-3}) as
\begin{equation}
M_{\rm MS}=\frac{\alpha^2}{4R}(X^2-T^2).
\label{eq:ms-mass}
\end{equation} 
It should be noted that the Misner-Sharp mass 
is given by $M_{\rm MS}=F/2$, where $F$ has been defined by Eq.(\ref{eq:ein4}).

\subsection{The case of $\alpha^2>1/2$}

The domain covered by the original chart is $0<T<X$. 
It is easily seen from Eqs.~(\ref{eq:metric-2}) and (\ref{eq:TX-R}) that 
$T=X$ is regular (see also Appendix \ref{ApRicci}). 
Thus we can analytically extend the region $0<T<X$ 
covered by the original chart (\ref{eq:metric-2}) over the region $0<T<\sqrt{2}\alpha X$ 
where the positivity of $R^2$ is guaranteed. 
This extension have also been discussed in Oshiro et al \cite{Oshiro}.
$R^2$ vanishes on $T=\sqrt{2}\alpha X$. 
Here it should be noted that $T=\sqrt{2}\alpha X$ is a spacetime singularity where  
the energy density $\rho$ and scalar polynomials diverge there: see 
Eq.~(\ref{eq:TX-rho}) and Appendix \ref{ApRicci}. 
This singularity is a timelike naked singularity.

\begin{figure}
\begin{center}
\includegraphics[width=0.4\textwidth]{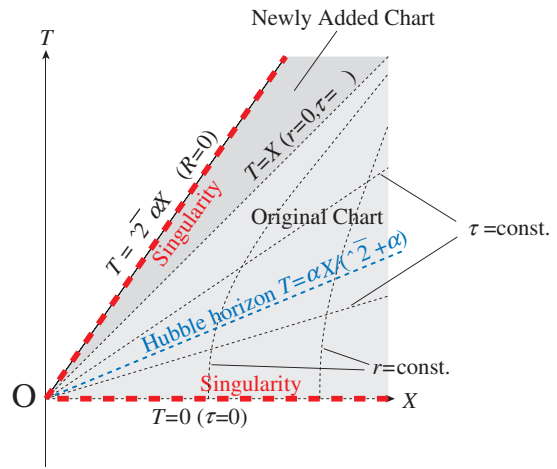}
\caption{\label{fg1}
The singularities and Hubble horizon of the analytically extended solution 
are depicted in $(T,X)$-plane. 
}
\end{center}
\end{figure}

\begin{figure}
\begin{center}
\includegraphics[width=0.4\textwidth]{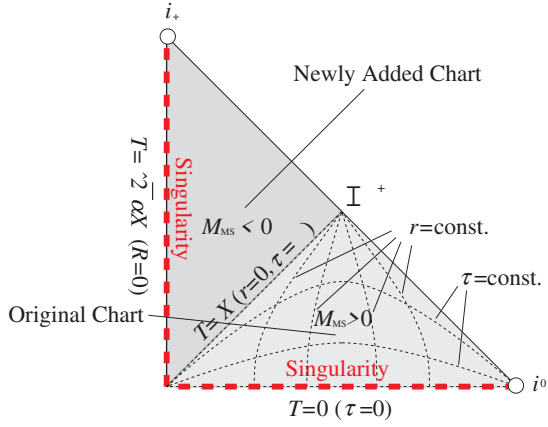}
\caption{\label{fg2}
The conformal diagram of analytically extended solution is depicted. 
This is similar to the upper half of the Minkowski spacetime. Note that the comoving lines, 
$r=$constant, enter into the future null infinity. This means that the elements of stiff fluid 
are accelerated to the speed of light asymptotically. 
}
\end{center}
\end{figure}

It is seen from Eq.~(\ref{eq:out-expansion}) that the expansion of outgoing null 
$\vartheta_{+}$ is positive everywhere in the maximally extended chart, 
whereas the expansion of ingoing null $\vartheta_{-}$ is negative for 
$\alpha X/(\sqrt{2}+\alpha)<T<\sqrt{2}\alpha$, vanishes on 
$T=\alpha X/(\sqrt{2}+\alpha)$, and is positive 
$0<T<\alpha X/(\sqrt{2}+\alpha)$. 
Thus, the region of $0<T<\alpha X/(\sqrt{2}+\alpha)$ is past trapped, 
$T=\alpha X/(\sqrt{2}+\alpha)$ is a past marginal surface, 
and $\alpha X/(\sqrt{2}+\alpha)<T<\sqrt{2}\alpha X$ is untrapped. 
The past marginal surface $T=\alpha X/(\sqrt{2}+\alpha)$ 
corresponds to the Hubble horizon. 
The singularities and the Hubble horizon in $(T,X)$-plane are depicted in 
Fig.~\ref{fg1}, while 
the conformal diagram is given in Fig.~\ref{fg2}. 
Here it should be noted that the Hubble horizon is spacelike; this is an important 
difference between the present solution and inflationary solutions.

We can easily see from Eq.~(\ref{eq:ms-mass}) that 
$M_{\rm MS}$ is positive for $0<T<X$, $M_{\rm MS}$ vanishes at $T=X$ ($r=0$ in 
original coordinate chart), and $M_{\rm MS}$ is negative for $T>X$. 
Further, $M_{\rm MS}=-\infty$ at $R=0$, i.e., 
$T\rightarrow \sqrt{2}\alpha X_{+0}$. 
The contribution of the scalar fields to $M_{\rm MS}$ is 
positive and thus $M_{\rm MS}$ is a monotonically increasing function 
with respect to $X$. Thus the negativity of Misner-Sharp mass 
in the region added by the extension comes from 
the negative infinite Misner-Sharp mass concentrated on the central 
timelike singularity. 
It seems difficult to get physical interpretation of this analytically 
extended solution, since the effect of the central naked singularity on its 
surrounding region must not be negligible. 

The scalar field now takes the following form, 
\begin{equation}
\phi=\frac{1}{\sqrt{2}}\ln\biggl|\frac{T\sqrt{2\alpha^2-1}}{\sqrt{2}\alpha X-T}\biggr|. 
\end{equation}
At the singularities $T=0$ and $T=\sqrt{2}\alpha X$, the scalar field diverges. Here, it is 
interesting to see the norm of the gradient of the scalar field. We have
\begin{equation}
g^{ab}\phi_{;a}\phi_{;b}=\frac{\alpha^2}{R^4}(T^2-X^2).
\end{equation}
We see from the above equation that $\phi_{;a}$ is timelike in the domain $0<T<X$, 
null on $T=X$ and spacelike in $X<T<\sqrt{2}\alpha X$. Therefore, 
this analytically extended solution is not equivalent to 
the system of a stiff perfect fluid.

\subsection{The case of $\alpha^2=1/2$}

In this case, from Eq.~(\ref{eq:TX-rho}), we find that 
the energy density $\rho$ diverges at $T=X$, and thus $T=X$ is a null spacetime 
singularity which is not naked, and its Misner-Sharp mass vanishes.  
As a result, the original chart is inextendible. 
Also in this case, $T=X/3$ is a past marginal surface   
which corresponds to the Hubble horizon. 
The singularities and the Hubble horizon in $(T,X)$-plane are depicted 
in Fig.~3. The conformal diagram is given in Fig.~4.
By contrast to the case of $\alpha^2>1/2$, 
comoving lines of constant $r$ enter into a future timelike infinity. 
There is a future null and future timelike and spacelike infinities.  

\begin{figure}
\begin{center}
\includegraphics[width=0.4\textwidth]{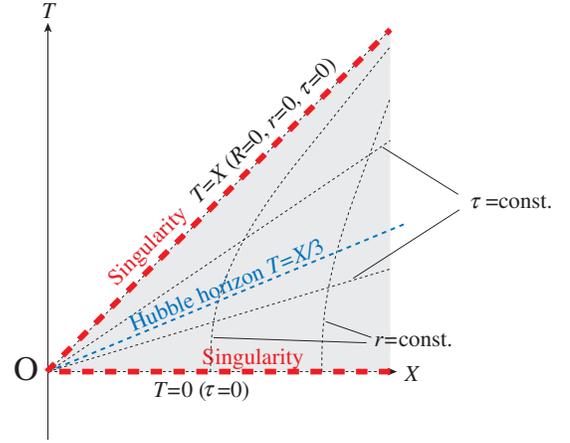}
\caption{\label{fg3}
The singularities and Hubble horizon for $\alpha^2=1/2$ 
are depicted in $(T,X)$-plane. There is a central null singularity which is not naked. 
}
\end{center}
\end{figure}

\begin{figure}
\begin{center}
\includegraphics[width=0.5\textwidth]{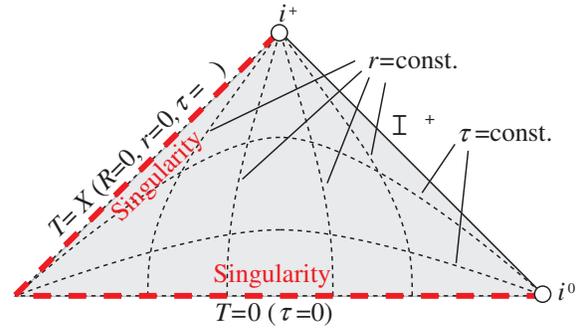}
\caption{\label{fg4}
The conformal diagram of analytically extended solution of $\alpha^2=1/2$ is depicted. 
The comoving lines, $r=$constant, enter into the timelike infinity. 
}
\end{center}
\end{figure}

\subsection{The case of $\alpha^2<1/2$}

As mentioned earlier, in this case, there are two kinds of 
singularities in the universe. Both of them are spacelike: one is the big bang 
and the other is the big crunch. It is remarkable 
that even though this is not a closed universe in the usual sense, 
the big crunch exists. 
There are the past and future marginal surfaces. Both of them correspond 
to the Hubble horizon, but the past one is in the expanding phase, whereas 
the future one is in the contracting phase. There is no null infinity. 
This type of solution is called the ``universal'' black hole by 
Carr and Gundlach\cite{Carr_Gundlach}. 
The same solution as this solution has also been obtained by Carr, 
Harada and Maeda\cite{CHM}. The singularities and past marginal surface are depicted 
in $(T,X)$-plane in Fig.~5. The conformal diagram is also given in Fig.~6.

\begin{figure}
\begin{center}
\includegraphics[width=0.4\textwidth]{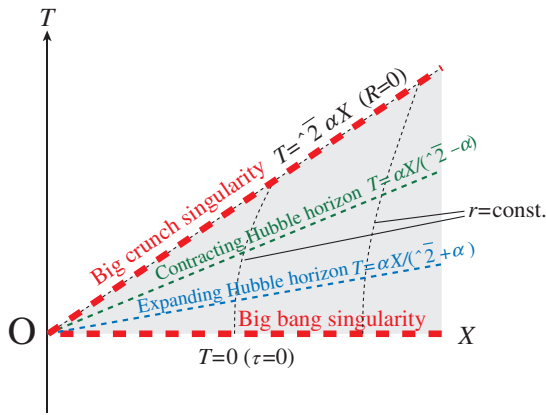}
\caption{\label{fg5}
The singularities and Hubble horizon for $0<\alpha^2<1/2$ 
are depicted in $(T,X)$-plane. 
}
\end{center}
\end{figure}

\begin{figure}
\begin{center}
\includegraphics[width=0.5\textwidth]{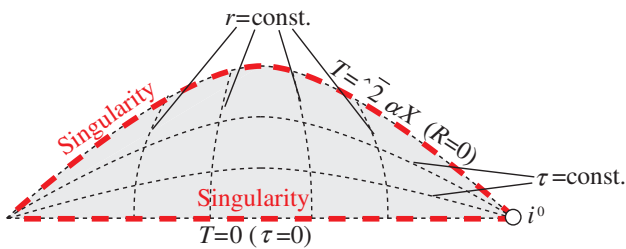}
\caption{\label{fg6}
The conformal diagram of analytically extended solution of $\alpha^2<1/2$ is depicted. 
There are both the big bang and big crunch singularities although the topology of 
$t=$constant hypersurface is ${\bf R}^3$. 
}
\end{center}
\end{figure}

\section{$C^1$ Extension for $\alpha^2>1/2$: spherical void solution}

Here, we consider another extension for the case of $\alpha^2>1/2$ 
without any serious singularities except for the big bang. 
For this purpose, it is very important to note that 
the Misner-Sharp mass vanishes on $T-X=0$, i.e.,  
$(\tau,r)=(\infty,0)$ in original chart. 
This implies that the extended region $T>X$ 
can be Minkowski spacetime  where the Misner-Sharp mass vanishes. 
In order to make clear whether such extension is possible, we consider 
the following Lorentz boost in the subspace $(T,X)$, 
\begin{equation}
T=\frac{\bar{T}+v\bar{X}}{\sqrt{1-v^2}}~~~~
{\rm and}~~~~X=\frac{\bar{X}+v\bar{T}}{\sqrt{1-v^2}},
\label{eq:boost}
\end{equation}
where
\begin{equation}
v=\frac{\sqrt{2}}{\alpha}-1.
\end{equation}
Then, we have the metric, for $T-X\leq0$, as
\begin{equation}
ds^2=-d\bar{T}^2+d\bar{X}^2+R^2d\Omega^2,
\end{equation}
where
\begin{equation}
R^2=\bar{X}^2-\frac{\alpha^2}{2}\left(\bar{X}-\bar{T}\right)^2.
\end{equation}
By the Lorentz boost (\ref{eq:boost}), the null hypersurface $T-X=0$ is mapped to 
$\bar{T}-\bar{X}=0$, and thus we have $R^2=\bar{X}^2$ at $(\tau,r)=(\infty,0)$. 
Further, we have, at $(\tau,r)=(\infty,0)$, or equivalently $\bar{T}-\bar{X}=0$, 
\begin{equation}
\partial_{\bar{T}}R^2=0~~~~{\rm and}~~~~\partial_{\bar{X}}R^2=2\bar{X}
\end{equation}
Thus, if we put, for $\bar{T}+\bar{X}<0$, the Minkowski spacetime 
\begin{equation}
ds^2=-d\bar{T}^2+d\bar{X}^2+\bar{X}^2d\Omega^2, \label{eq:M-metric}
\end{equation}
we can easily see that this is a $C^1$ extension of the original solution. 
Note that the Ricci tensor of this solution does not vanish on $T=X$ and hence 
the second order derivatives of the metric tensor are discontinuous. 
Also this extension have been discussed in Oshiro et al \cite{Oshiro}.

\begin{figure}
\begin{center}
\includegraphics[width=0.4\textwidth]{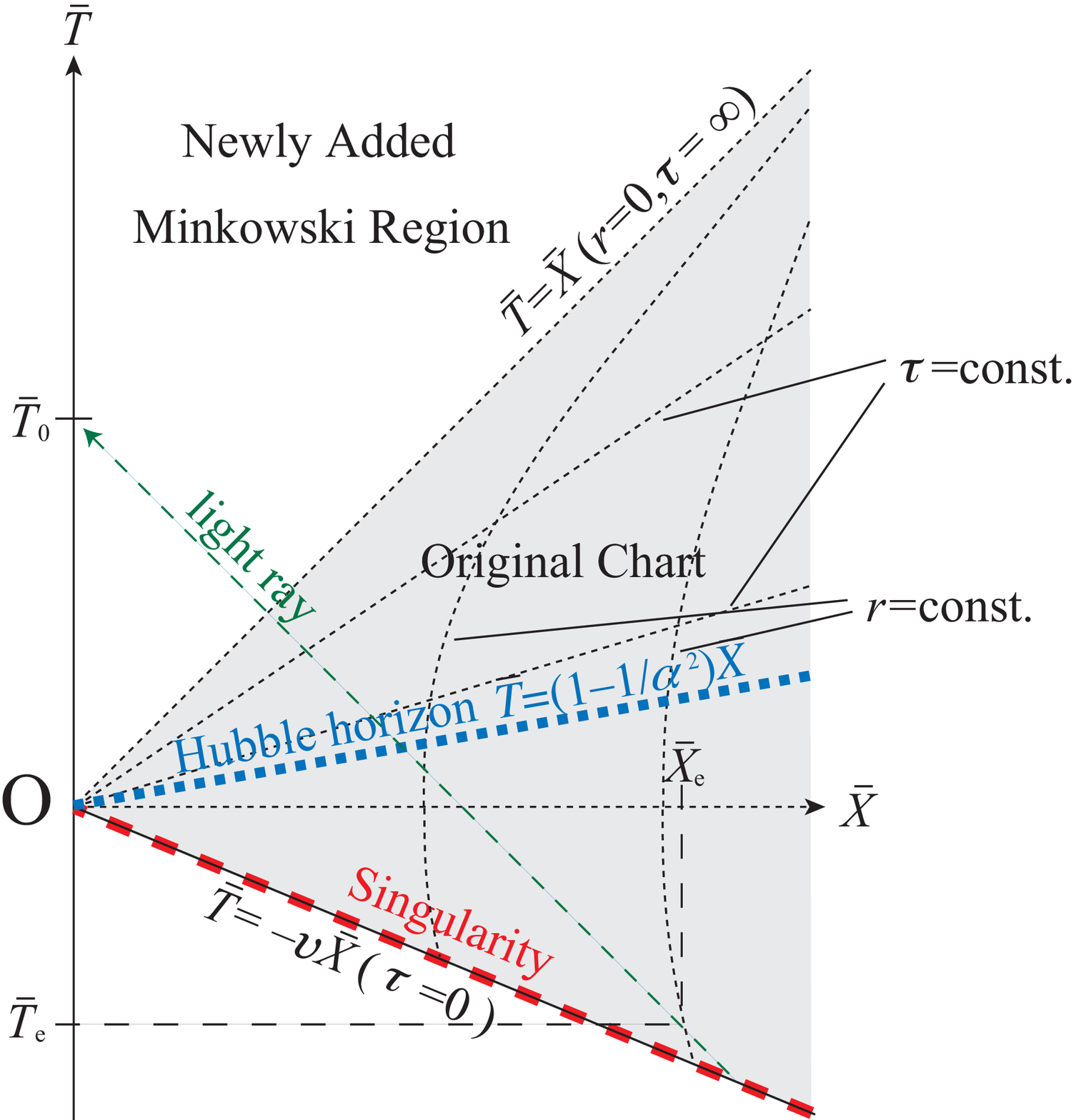}
\caption{\label{fg7}
In the case of $C^1$-extended solution, 
the big bang singularity, cosmological horizon and light ray are depicted in $(T,X)$-plane. 
}
\end{center}
\end{figure}

\begin{figure}
\begin{center}
\includegraphics[width=0.5\textwidth]{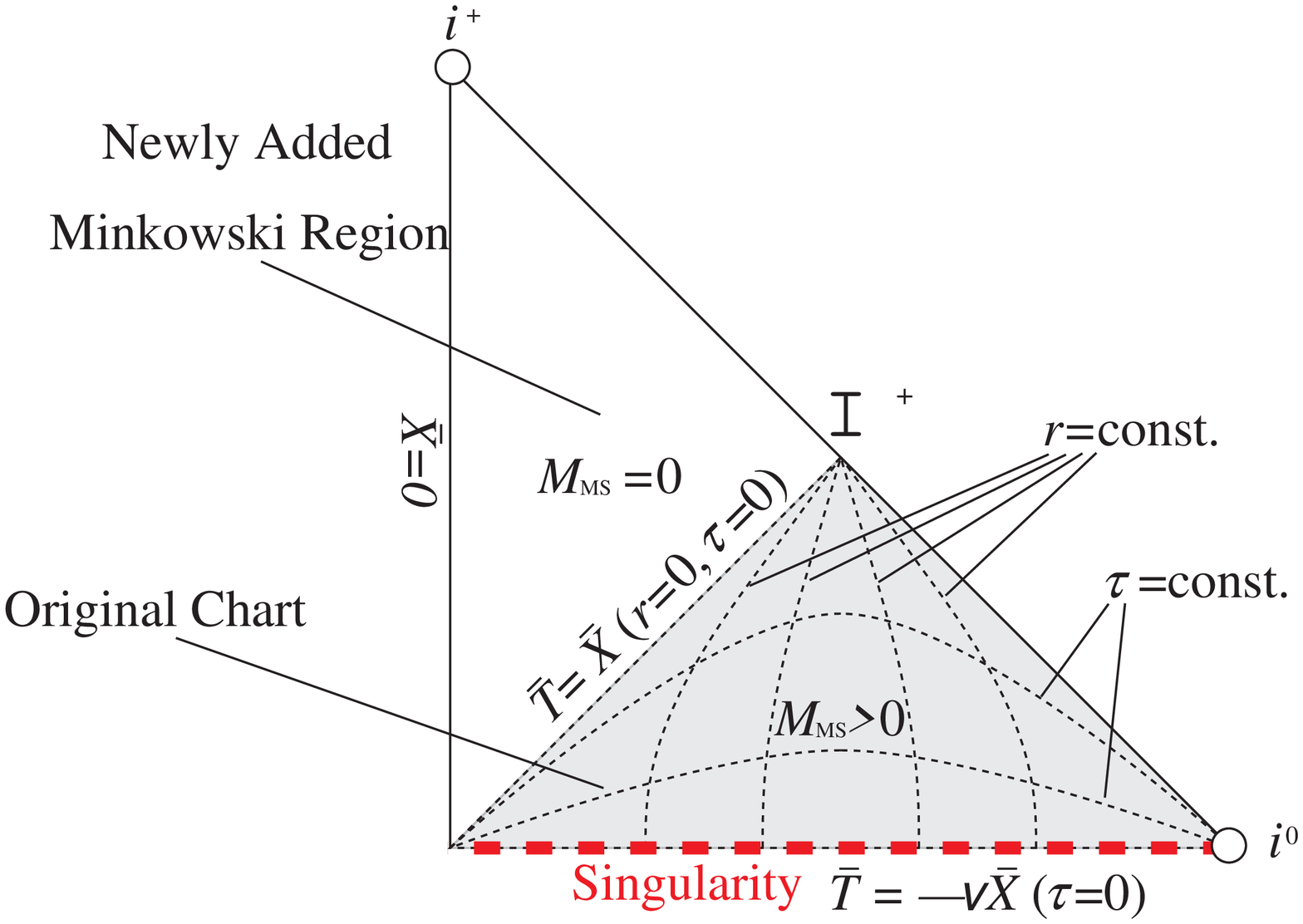}
\caption{\label{fg8}
The conformal diagram of $C^1$ maximally 
extended solution is depicted. This solution describes 
an expanding spherical void in the universe filled with 
a massless scalar-field. 
}
\end{center}
\end{figure}

The region newly added by this extension is regarded as an 
expanding void. 
The big bang singularity and cosmological horizon of this $C^1$ extended spacetime 
are depicted in $(\bar{T},X)$-plane in Fig.\ref{fg7}. 
The conformal diagram of this $C^1$ extended spacetime is depicted 
in Fig.\ref{fg8}. 

This universe model has a few very peculiar properties. Since the gradient of the 
scalar field is null on $\bar{T}=\bar{X}$, this solution cannot be regarded as 
the universe filled with a stiff perfect fluid only. However, 
we can regard the matter field in $\bar{T}<\bar{X}$ as the stiff perfect fluid. 
The world lines of fluid elements 
are the curves with constant $r$. These 
are expressed in $(\bar{T},\bar{X})$-coordinate system by hyperbolic curves
\begin{equation}
\bar{X}^2-\bar{T}^2=r^2.
\end{equation}
We see from the above equation that the world lines of fluid elements 
become asymptotically null. This fact might be rather surprising, since 
the fluid elements are accelerated outward even though there 
is a positive gravitational
mass (Misner-Sharp mass) inside the mass shell on which 
the fluid elements stay.

The outward acceleration causes the temporal growth of the redshift of 
the light ray emitted from a fluid element in $\bar{T}<\bar{X}$ 
to the observer at the symmetry center $\bar{X}=0$. (see Fig.\ref{fg8}.) 
The ingoing null geodesics are given by
\begin{equation}
\bar{T}=\omega_0\lambda+\bar{T}_0~~~{\rm and}~~~\bar{X}=-\omega_0\lambda,
\end{equation}
where $\lambda$ is the affine parameter, $\omega_0$ and $\bar{T}_0$ are positive  
constants which correspond to the angular frequency detected by the observer 
at $(\bar{T},\bar{X})=(\bar{T}_0,0)$. The big bang singularity 
is given by $\bar{T}=-v\bar{X}$ in this coordinate system, 
and thus the affine parameter 
takes the values in $-T_0/\sqrt{2}\omega_0(1-v)<\lambda\leq 0$. 
The tangent vector of this null geodesic is 
\begin{equation}
k^\mu=\left(\omega_0,-\omega_0,0,0\right).
\end{equation}
The 4-velocity of the fluid element labeled by $r$ is given by
\begin{equation}
u^\mu=\frac{1}{r}\left(\sqrt{\bar{T}^2+r^2},\bar{T},0,0\right).
\end{equation}
The light ray detected by the observer at $(\bar{T},\bar{X})=(\bar{T}_0,0)$ should be emitted 
from the fluid element of $r$ at
\begin{equation}
(\bar{T},\bar{X})=(\bar{T}_{\rm e},\bar{X}_{\rm e}):=
\left(\frac{\bar{T}_0^2-r^2}{2\bar{T}_0},
\frac{\bar{T}_0^2+r^2}{2\bar{T}_0}\right).
\end{equation}
Thus the redshift $z$ of this light ray is given by
\begin{equation}
z=\frac{-k^a u_a|_{(\bar{T},\bar{X})=(\bar{T}_{\rm e},\bar{X}_{\rm e})}}{\omega_0}-1
=\frac{\bar{T}_0-r}{r}.
\end{equation}
The important and peculiar feature is that the redshift $z$ is monotonically decreasing 
function of $r$. 
For a fluid element of $\bar{T}_0<r$, the redshift $z$ is negative, i.e., the 
light suffers the blueshift. The fluid element of $\bar{T}_0<r$ emits the 
light ray at $\bar{T}=\bar{T}_{\rm e}<0$, and at this moment, the radial component 
of the 4-velocity is negative. Thus the reason of the 
blueshift for $\bar{T}_0<r$ is recognized as a result of 
the ingoing motion of the fluid element. By contrast, if $\bar{T}_0>r$, i.e., 
$\bar{T}_{\rm e}>0$, the redshift $z$ is positive. 
If $\bar{T}_0=r$, i.e., $\bar{T}_{\rm e}=0$, the redshift vanishes. 
The temporal variation rate of $z$ is 
\begin{equation}
\frac{dz}{d\bar{T}_0}=\frac{1}{r}>0.
\end{equation}
The redshift temporally increases. This is also a very different property 
from the dust-filled inhomogeneous universe model\cite{Uzan,Yoo}. 

Since the cosmological horizon $\bar{T}=(1-1/\alpha^2)\bar{X}$, 
or equivalently, $T=\alpha X/(\sqrt{2}+\alpha)$, is spacelike, the fluid elements 
necessarily enter into the inside of the cosmological horizon from its outside. 
Thus even though the fluid elements are accelerated outward, the 
inflation does not occur in usual sense. 
(We note that in the inflationary period, comoving world lines go 
outside the cosmological 
horizon from its inside, and such behaviors of comoving world lines are 
essential to resolve the horizon problem.)

Finally, we investigate the distance-redshift relation. Observationally useful 
distance is the luminosity distance $d_{\rm L}$\cite{Weinberg} 
which is given by
\begin{eqnarray}
d_{\rm L}&=&(1+z)^2R|_{(\bar{T},\bar{X})=(\bar{T}_{\rm e},\bar{X}_{\rm e})} \nonumber \\
&=&\frac{\bar{T}_0}{2}\sqrt{(1+z)^4+2(1+z)^2-2\alpha^2+1}.
\end{eqnarray}
Since $d_{\rm L}$ is monotonically increasing function of $z$, it is monotonically decreasing 
function with respect to $r$. 
The redshift of the light ray from the big bang $z=z_{\rm b}$ is given by
\begin{equation}
z_{\rm b}=\sqrt{\sqrt{2}\alpha-1}-1.
\end{equation}
We can easily see that the luminosity distance of the big bang vanishes. Since 
the light ray emitted from a comoving source at 
the big bang suffers finite blueshift, the flux from the big bang 
${\cal F}_{\rm b}=L_{\rm b}/4\pi d_{\rm L}^2$ diverges at the observer, 
if the luminosity of the big bang singularity $L_{\rm b}$ is finite. 
This feature implies that comoving sources of radiation 
behave as white holes at this big bang singularity\cite{Novikov,Neeman}. 
In the limit $r\rightarrow0_{+}$, $z\rightarrow\infty$ and thus 
$d_{\rm L}$ of the boundary of void $r=0$ is infinite. 
It is hard to observe the vicinity of the boundary of the 
void, while it is easy to observe the vicinity of the big bang, in this universe.

\section{Summary and Discussion}

We have rederived and investigated here the solution obtained 
by Roberts, which is a spherically symmetric but inhomogeneous universe 
filled with a massless scalar field minimally coupled to gravity, 
from a cosmological perspective. 
The solutions  obtained  compose a one-parameter family, which is 
divided into three distinct  classes. 
The first class consists of solutions with only one spacelike singularity in the 
comoving chart. The second class consists of solutions 
with two singularities which are null and spacelike, respectively. 
The third class consists of solutions with two spacelike singularities 
which correspond to the big bang and big crunch, respectively. 

In the case of the first class, the comoving chart does not 
cover the whole spacetime. 
Hence, we constructed two maximally extended solutions 
from  the solution in this class. 
The analytic extension leads to the solution which contains 
a timelike singularity at the symmetry center and thus seems to be 
unphysical. By another extension, we obtained a solution which 
has no singularity other than the big bang but one that contains 
a spherical void. 
The later one has very peculiar but interesting properties. 
If the gradient of the massless scalar field is timelike, this is equivalent 
to the stiff perfect fluid case. Then in the region where the scalar field has 
a timelike norm, we can naturally define fluid elements and consider 
their motions. In this solution, the fluid elements move outwards, 
and further, their outward speeds are accelerated, even though 
the strong energy condition holds. This is a feature which 
is significantly different  from the homogeneous and isotropic universe 
models filled with the matter which satisfies the strong energy condition, 
and also from spherically symmetric but inhomogeneous 
universe filled with the dust matter. 

The physical reason of the outward acceleration of the universe
that we have deduced here appears to lie essentially in the inherent 
physical nature itself of the massless scalar field. 
To try to understand this better, we note that we can construct a similar 
and somewhat parallel situation in a Minkowski background. 
Using the same coordinate system as Eq.~(\ref{eq:M-metric}), 
let us consider a following solution, 
\begin{equation}
\phi(\bar{T},\bar{X})=
\frac{A}{\bar{X}}(\bar{X}-\bar{T})^2
\theta_{\rm H}(\bar{X}-\bar{T})
\end{equation}
where $A$ is constant and $\theta_{\rm H}$ is the Heaviside's step function. 
We focus on the domain of $\bar{T}>0$, since $\phi$ is everywhere finite for 
$\bar{T}\geq 0$, but is infinite at $\bar{X}=0$ for $T<0$. Then we have
\begin{equation}
g^{ab}\phi_{;a}\phi_{;b}=-\frac{A^2}{\bar{X}^4}(\bar{X}-\bar{T})^3(3\bar{X}+\bar{T})
\theta_{\rm H}(\bar{X}-\bar{T}).
\end{equation}
We can easily see from the above equation that $\phi_{;a}$ is timelike for 
$\bar{T}<\bar{X}$, whereas it is null at $\bar{T}=\bar{X}$. 
Further, we see that in the limit of $\bar{T}\rightarrow\infty$ with 
$\bar{U}:=\bar{T}-\bar{X}$ fixed, the norm of $\phi_{;a}$ becomes 
\begin{eqnarray}
g^{ab}\phi_{;a}\phi_{;b}\rightarrow
\lim_{\bar{T}\rightarrow\infty}
\frac{A^2}{(\bar{T}-\bar{U})^4}\bar{U}^3(4\bar{T}+3\bar{U})
=0.
\end{eqnarray}
The above results shows that $\phi_{;a}$ becomes null asymptotically. 
This behavior is very similar to the present solution. 
This seems to indicate that the outward acceleration comes 
from the inherent nature of the massless scalar field. As it turns out, 
in the present solution, the gravity produced by the 
scalar field itself is too small to decelerate the outward motion 
of the scalar field, even though the total (Misner-Sharp) 
mass for the universe is infinite.  

\section*{Acknowledgments}

KN are grateful to H.~Ishihara and colleagues in the research group of 
elementary particle physics group and of gravity group at Osaka City University 
for their useful and helpful discussion and criticism. 
KN also thanks to H. Abe, Y. Morisawa, Y. Takamori for their careful check of 
the present exact solution. This work is supported by the Grant-in-Aid for Scientific 
Research (No.21540276).

\appendix
\section{On the function $f(r)$}\label{ApA}

For notational simplicity, we define the following quantities
\begin{equation}
\chi:=r^2f(r)~~~~{\rm and}~~~~u:=g(t)+p(r),
\end{equation}
where $f$ and $p$ are defined by Eqs.~(\ref{kg}) and (\ref{eq:p-def}), respectively. 
Then, Eq.~(\ref{eq:ein4-2}) is written in the form
\begin{equation}
H+1=\frac{u^2}{2}+\frac{hu}{\chi} 
+\left(\frac{\chi'}{u}+\frac{h'}{u^2}\right)^2.
\end{equation}
By differentiating the above equation with respect to $r$, we have
\begin{equation}
c_6u^6+c_5u^5+c_4u^3+c_3u^2+c_2u+c_1=0, \label{eq:poly}
\end{equation}
where
\begin{eqnarray}
c_1&=&-4{h'}^2p', \\
c_2&=&2\left(h'h''-3\chi'h'p'\right), \\
c_3&=&2\left[\left(\chi'h'\right)'-{\chi'}^2p'\right], \\
c_4&=&2\chi'\chi'', \\
c_5&=&\frac{hp'}{\chi},\\
c_6&=&\left(p+\frac{h}{\chi}\right)'. 
\end{eqnarray}
Here note that all of the coefficients $c_n$ do not depend on $t$. This implies that 
if there is a non-trivial real root $u=U$ of Eq.~(\ref{eq:poly}), $U$ must  
depend on only $r$. If so, we have
\begin{equation}
g(t)=U(r)-p(r).
\end{equation}
The above equation holds only if both of $g$ and $U-p$ are 
constants. Thus, there are two possibilities; 
one is $g(t)=$constant, and the other is that all of the coefficients 
$c_n$ ($n=1,..,6$) identically 
vanish so that Eq.~(\ref{eq:poly}) is trivial. 
In the latter case, $g$ may depend on $t$. 

In the former case, without loss of generality, we can put $u=p$. 
Then, by definition of $p$, we find 
\begin{equation}
\chi=-\frac{h'}{p'}. \label{eq:chi-def-2}
\end{equation}
Further, substituting $\dot{g}=0$ into Eq.~(\ref{eq:g-dot}), we have $H=0$. 
By using this fact and Eq.~(\ref{eq:chi-def-2}), 
$\chi$ can be eliminated from Eq.~(\ref{eq:poly}), and we have
\begin{equation}
\left(p''-\frac{h''}{h'}p'+\frac{{p'}^2}{p}\right)^2=\frac{p^2{p'}^4}{{h'}^2}
\left(1+\frac{hp}{h'}p'-\frac{p^2}{2}\right).
\end{equation}
The above differential equation gives a relation between 
$h(r)$ and $p(r)$.
The above equation, however, corresponds to the static
case, in which we are not interested here in the present
work. The static solutions will be discussed elsewhere.

As for the dynamical case, the conditions $c_n=0$ give 
differential 
equations for $\chi$ and $h$. From $c_6=0$, we have
\begin{equation}
\frac{h}{\chi^2}\chi'=0.
\end{equation}
Thus, we have $h=0$ or $\chi'=0$. From $c_5=0$, we have 
$hh'=0$, and thus 
\begin{equation}
h={\rm constant}. \label{eq:h-sol}
\end{equation}
From $c_4=0$, we have $\chi'=$constant, and thus
\begin{equation}
\chi=\alpha r+ \beta, \label{eq:chi-sol}
\end{equation}
where $\alpha$ and $\beta$ are integration constants. 
Eqs.~(\ref{eq:h-sol}) and(\ref{eq:chi-sol}) lead to $c_1=c_2=c_3=0$. 
Since we assume that $r=0$ is a non-singular point at least initially, 
$R$ vanishes at $r=0$. Thus, we can see from Eq.~(\ref{kg}) 
that $\beta$ should vanish. Further, it is seen from Eq.~(\ref{eq:ein4}) 
that $F$ should vanish at the regular origin $r=0$. Hence, 
from Eq.~(\ref{eq:F-sol}), $h$ should vanish at $r=0$, 
and thus, we have $h=0$.

\section{Misner-Sharp Mass}\label{ApMS}

In general, the metric of the spherically symmetric spacetime is given by
\begin{equation}
ds^2=A^2(t,r)dt^2-B^2(t,r)dr^2-R^2(t,r)d\Omega^2.
\end{equation}
In this coordinate system, the Misner-Sharp mass is given by\cite{Misner}  
\begin{equation}
M_{\rm MS}=\frac{R}{2}\left(1+\frac{R^2}{2}\vartheta_{+}\vartheta_{-}\right),
\label{mass-def}
\end{equation}
where
\begin{equation}
\vartheta_\pm =\frac{1}{\sqrt{2}}\left(\frac{1}{A}\partial_t\pm\frac{1}{B}\partial_r\right)
\ln R^2.
\end{equation}
Here note that $\vartheta_{+}$ is the expansion of the outgoing radial null, while 
$\vartheta_{-}$ is the expansion of the ingoing null. 

\section{Ricci tensor}\label{ApRicci}

Non-vanishing components of the Ricci tensors with respect to the analytically 
extended chart are
\begin{eqnarray}
Ric\left(\frac{\partial}{\partial T}, \frac{\partial}{\partial T}\right)&=&
\frac{\alpha^2 X^2}{R^4}, \\
Ric\left(\frac{\partial}{\partial T}, \frac{\partial}{\partial X}\right)&=&
-\frac{\alpha^2 TX}{R^4}, \\
Ric\left(\frac{\partial}{\partial X}, \frac{\partial}{\partial X}\right)&=&
\frac{\alpha^2T^2}{R^4}.
\end{eqnarray}
The Ricci scalar is then given by
\begin{equation}
{\rm tr}Ric=\frac{\alpha^2}{R^4}(T^2-X^2).
\end{equation}
Since $R$ vanishes on $T=\sqrt{2}\alpha X$, 
the Ricci scalar ${\rm tr}Ric$ diverges there.

\end{document}